# Hyperbolic Weyl point in reciprocal chiral metamaterial

Meng Xiao[1], Qian Lin[2] and Shanhui Fan[1,*]

[1]*Department of Electrical Engineering, and Ginzton Laboratory, Stanford University, Stanford, California 94305, USA*

[2]*Department of Applied Physics, Stanford University, Stanford, California 94305, USA*

[*]Corresponding E-mail: shanhui@stanford.edu

**Abstract**:

We report the existence of Weyl points in a class of non-central symmetric metamaterials, which has time reversal symmetry, but does not have inversion symmetry due to chiral coupling between electric and magnetic fields. This class of metamaterial exhibits either type-I or type-II Weyl points depending on its non-local response. We also provide a physical realization of such metamaterial consisting of an array of metal wires in the shape of elliptical helixes which exhibits type-II Weyl points.

Weyl point is a topologically singular point in three-dimensional wave vector space. Initially studied in electronic systems [1-10], in recent years the physics of Weyl point has also been studied in classical wave systems such as electromagnetic[11-16] and acoustic systems[17,18] Compared with the electronic systems, in classical wave systems one has more flexible control of the geometry, and the coherent transport properties are easier to measure due to the lack of electron-electron scattering. Therefore the classical wave systems complements the electronic systems in the demonstrations of various signatures of the Weyl points. In addition to the fundamental interest, the use of Weyl points may also lead to new capabilities for manipulating the propagation of electromagnetic and acoustic waves.

Recently it was noted that here are two types of Weyl points, type-I and type-II, both are topologically nontrivial but exhibit very different physical properties. [10] Up to now, there has not been any exploration of type-II Weyl points in the electromagnetic systems. Here we report the existence of Weyl points in a class of non-central symmetric electromagnetic metamaterial which preserves time reversal symmetry. The breaking of inversion symmetry is achieved through chiral coupling between electric and magnetic fields. This class of metamaterial exhibits either type-I or type-II Weyl points depending on its non-local response.[19,20] We also provide a physical realization of such metamaterial consisting of an array of metal wires in the shape of elliptical helixes which exhibits type-II Weyl points.

The simplest model Hamiltonian which exhibits both types of Weyl points has the form

$H = \tau k_z \mathbf{I} + k_x \sigma_x + k_y \sigma_y + k_z \sigma_z$, [10] where $k_x$, $k_y$ and $k_z$ are the wave vectors along the $x$, $y$ and $z$ axes respectively, $\sigma_x$, $\sigma_y$ and $\sigma_z$ are the Pauli matrices, $\mathbf{I}$ is a $2 \times 2$ identity matrix and $\tau$ is a real constant. In this model, the group velocities of the two bands along the $z$ direction are $\tau + 1$ and $\tau - 1$ respectively, whereas along either the $x$ or $y$ directions, the group velocities for the two bands are $+1$ and $-1$, respectively. When $\tau \in (-1,1)$, the Hamiltonian exhibits a type-I Weyl point at $\vec{k} = 0$. The constant frequency surface around $\vec{k} = 0$ is an ellipsoid. Thus one can also refer to such a Weyl point as an elliptical Weyl point. On the other hand, when $\tau \notin [-1,1]$, the model Hamiltonian exhibits a type-II Weyl point. The constant frequency surface is now a hyperboloid. Therefore, we will also refer to such type-II Weyl point as a hyperbolic Weyl point. Both types of Weyl points are topologically non-trivial, but they exhibit very different physical properties. For example, while the density of state is zero at a type-I Weyl point, at a type-II Weyl point, the density of state in fact diverges for this model, similar to hyperbolic metamaterials [21] and can be rather large in the physical realization of this model. [10]

Here we report a class of electromagnetic metamaterial that exhibits type-II Weyl points in the homogenization limit. There have been significant recent interests seeking to study the physics of Weyl points in electromagnetic[12-16] and acoustic systems[17,18]. However, none of the previous works have reported the observation of a type-II Weyl point. In addition, most of the previous works concerning Weyl point in classical wave systems utilize either photonic or phononic crystals, where the periodicity plays a significant role. The exploration of Weyl point in metamaterials as described by homogeneous effective material parameters is of fundamental interest since the wavevector space of such meta-material is non-compact, which is in contrast with the wavevector space of periodic systems which is always topologically compact.[22] As a result, there is significant qualitative difference between the topological properties of the bands for meta-materials and photonic crystals as we will discuss in this paper.

Since the Berry curvature vanishes in systems that possess both time reversal and inversion symmetries, [11] one needs to break either time-reversal or inversion symmetries to construct a topologically non-trivial object such as a Weyl point. The existence of Weyl points in homogeneous material without time-reversal symmetry have been reported very recently.[16] However, the system in ref. 16 requires magneto-optical effects under a large static magnetic field, which is challenging to demonstrate experimentally. In this work, we show that Weyl points can be achieved in reciprocal meta-materials where the inversion symmetry is broken with chiral coupling between electric and magnetic fields. And we provide a physical construction of such meta-material with an array of helix structure that should be relatively straightforward to construct experimentally.

**Weyl point in homogeneous material**

To construct a meta-material supporting a Weyl point we start with a nonmagnetic ($\mu_r = 1$) anisotropic homogeneous material. The relative permittivity of this material is given by $\ddot{\varepsilon}_r = \mathrm{diag}(\varepsilon_t, \varepsilon_t, \varepsilon_z)$, where $\varepsilon_z = 1 - \omega_p^2 / \omega^2$ possesses Drude's dispersion. Along the $z$-direction, the system supports three propagating modes, one of which is longitudinal and the other two are transverse. The longitudinal mode is given by $\varepsilon_z = 0$, and hence the dispersion relation of this mode is flat with frequency of the mode fixed at $w = w_p$ independent of the wavevector $k_z$ along the z direction. The two transverse modes, having their electric field in the *xy*-plane, are influenced only by the permittivity of the transverse directions, *i.e.*, $\varepsilon_t$. Since in this case the system is isotropic along the transverse directions, these two transverse modes are degenerate with their dispersion relation given by $\omega = ck_z / \sqrt{\varepsilon_t}$, where $c$ is the speed of light in vacuum. The dispersion relations of these three modes are shown in Fig. 1(a). They intersect at $k_z = k_c = \omega_p \sqrt{\varepsilon_t} / c$. Therefore, the photon bands are three-fold degenerate at $k = (0,0,k_c)$. Fig. 1(b) shows the dispersion relation in the $k_x - k_y$ plane at $k_z = k_c$. Two of the three bands form a conical dispersion around the origin. There is also an additional band, which is quadratic in $k_x$ and $k_y$, passing through the conical point.

Since a Weyl point is two-fold degenerate, we first need to break this triple degeneracy. Here we introduce anisotropy into the relative permittivity by setting

$$\ddot{\varepsilon}_r = \text{diag}\{\varepsilon_x, \varepsilon_y, \varepsilon_z\}, \tag{1}$$

where without loss of generality we assume $e_x > e_y$. Along the *z*-axis, the dispersion relation of the two transverse modes become (see Supplemental Material S-I)

$$\begin{cases} \omega = ck_z / \sqrt{\varepsilon_x} \\ \omega = ck_z / \sqrt{\varepsilon_y} \end{cases} \tag{2}$$

as shown schematically in Fig. 1(c). For subsequent discussion we refer to these two bands as bands 1 and 2, respectively. Band 1 thus has a lower frequency compared with band 2 at a given $k_z$. We refer to the longitudinal mode as band 3. By introducing the anisotropy, the triply degenerate point in the isotropic case now breaks into two linear crossing points. Consider the crossing point between bands 1 and 3. We use $k_c$ to represent the value of $k_z$ at this crossing point. Fig. 1(d) shows the dispersion relation in the $k_x - k_y$ plane at $k_z = k_c$. The dispersion is linear along the $x$ direction while quadratic along the $y$ direction. While we have created a two-fold degenerate crossing point, this point is not a Weyl point. (The dispersion of Weyl point is linear along all the directions) This is expected, since the structure has both time-reversal and inversion symmetry.

We break the inversion symmetry by introducing a chiral coupling term between the z-components of the electric and magnetic fields. [23,24] The constitutive relation becomes

$$\begin{pmatrix} \vec{D} \\ \vec{B} \end{pmatrix} = \begin{pmatrix} \varepsilon_0 \vec{\vec{\varepsilon}}_r & i\sqrt{\varepsilon_0 \mu_0} \vec{\vec{\gamma}} \\ -i\sqrt{\varepsilon_0 \mu_0} \vec{\vec{\gamma}} & \mu_0 \end{pmatrix} \begin{pmatrix} \vec{E} \\ \vec{H} \end{pmatrix}, \qquad (3)$$

where $\varepsilon_0$ and $\mu_0$ are the permittivity and permeability of the vacuum, respectively, $\vec{\vec{\varepsilon}}_r$ is as defined above in Eq. (1) and the only non-vanishing component of $\vec{\vec{\gamma}}$ is $\gamma_{zz}$. For a time-reversal symmetric system, $\vec{\vec{\varepsilon}}_r$ and $\gamma_{zz}$ are both real [25]. Consider again modes propagating along the z-direction, compared with the case where $g_{zz} = 0$, introducing $\gamma_{zz}$ doesn't affect the dispersions of the two transverse modes of bands 1 and 2. The only difference is that the dispersion of the longitudinal mode (i.e. band 3) is now determined by (see Supplemental Material S-I)

$$\varepsilon_z - \gamma_{zz}^2 = 1 - \omega_p^2/\omega^2 - \gamma_{zz}^2 = 0. \qquad (4)$$

From now on, we denote the frequency at which Eq. (4) is satisfied as $\omega_p'$. The dispersion relations for the three modes along the $z$ direction are shown schematically in Fig. 1(e). Same as before, we use $k_c$ to denote the value of the wave vector of the crossing point between bands 1 and 3. In Fig. 1(f), we show the dispersion in the $k_x - k_y$ plane at $k_z = k_c$. The dispersion becomes linear along all the directions indicating the crossing point is a Weyl point. (Plots of dispersion relations along other directions can be found in Supplemental Material Figs. S1(a)-(c).) Comparing Figs. 1(d) and 1(f), we see that the introduction of the chiral coupling drastically changes the dispersion relation. In Fig. 1(d), without the chiral coupling, the dispersion is quadratic along the $y$ direction, i.e., $\omega - \omega'_p \propto k_y^2$ around the crossing point. Whereas with the chiral coupling the dispersion becomes $\omega - \omega'_p \propto \gamma_{zz} k_y$, as shown in Fig. 1(f) (Detailed derivation of these results can be seen in Supplemental Material S-I) The other crossing point with a smaller $k_z$ value in the positive half $k_z$ space at $k_z = \sqrt{\mu \varepsilon_y} \omega'_p / c$ is also a Weyl point. Another two Weyl points are located in the negative $k_z$ space with $k_z = -\sqrt{\mu \varepsilon_x} \omega'_p / c$ and $k_z = -\sqrt{\mu \varepsilon_y} \omega'_p / c$, respectively, as required by time-reversal symmetry.

There exist many other choices of effective parameters for which the systems also exhibit Weyl points. For example, in real material, $\gamma_{zz}$ can be dispersive. (See Supplemental Material S-IV). Whether $\gamma_{zz}$ is dispersive or not does not affect the existence of the Weyl point. As another example, one can have Weyl points in systems that are isotropic in the $x$-$y$ plane, but have chiral coupling on all three directions. (See Supplemental Material S-V) The results here point to substantial opportunities for exploring Weyl point physics in a wide range of reciprocal chiral meta-materials.

**Charge of Weyl Points**

To further confirm that the crossing point in Fig. 1(e) is indeed a Weyl point and also to identify its topological charge, we numerically calculate the Berry phase over a closed spherical surface surrounding the crossing point in the wavevector space. The basic idea here is to compute the Berry phases of all the bands in the system, as the wavevectors vary around the circle at the intersection of a $k_x - k_y$ plane with the spherical surface, and to track the evolution of the Berry phase thus calculated as a function of $k_z$ as we change the $k_z$ from above to below $k_c$. This corresponds to a variation from $\theta = 0$ to $\theta = \pi$, as shown Fig. 3(a).[10] In general, as $\theta$ varies from 0 to $\pi$, the phases thus calculated can only vary from 0 to $2m\pi$, where $m$ is an integer. A monopole in the wavevector space corresponds to $m \neq 0$, with $m$ here being the topological charge, which is a quantity that is robust to parameter variation. The results of such calculation, for the crossing point between bands 1 and 3, are shown in Fig. 2(b), where black, red and blue lines represent the Berry phase of the lower, middle and upper bands, respectively. (The details of this calculation can be found in the Supplemental Material S-III). We see that the Berry phase of the lower (middle) band changes continuously from $2\pi$ to 0 (0 to $2\pi$) indicating that the charge of this crossing point is $-1$. Therefore, this is indeed a Weyl point. Meanwhile the Berry phase of the upper band is almost zero as there is no Weyl point on the upper band inside this sphere and the radius of the sphere is quite small. Similar calculation can also be done surrounding the other crossing point between band 2 and band 3, and the charge is also $-1$. Due to the time-reversal symmetry, the charges of the two Weyl points in the negative $k_z$ plane are also $-1$. We do not observe any other Weyl points. Therefore, the total charge of the Weyl points in this system is not zero. From the result above we see that the topological behavior of a uniform system can be distinctively different from that of a periodic system, such as the photonic and phononic crystal systems that are typically used for the study of Weyl point physics. For a periodic system, the wavevector space is compact. (For a cubic lattice, for example, the wavevector space is a 3-torus.) Consequently, the total charge of all the Weyl points inside the first Brillouin zone should vanish.[26] On the other hand, for a uniform system, the wavevector space is not compact, and there is no longer such a constraint on the total charge.

**Type of Weyl Points and Relation with Nonlocal Effect**

In the system above, one of the two bands forming the Weyl point is flat along the $z$-direction. The system therefore lies at the boundary that separates the phase spaces of systems exhibiting type-I and type-II Weyl points. Starting from the system discussed above, one can generate either type-I or II Weyl point using non-local effect that is widely known in meta-materials. [19,20] As an illustration, we assume a non-local $\varepsilon_z$ that depends only on $k_z$:

$$\varepsilon_z = 1 - \omega_p'^2 / \omega^2 + \gamma_{zz}^2 + \alpha k_z^2, \tag{5}$$

where we do a Taylor expansion in terms of $k_z$ and keep the lowest order in $k_z$ as allowed by time-reversal symmetry. The inclusion of higher order terms in the Taylor expansion typically only serves to shift the frequency of the Weyl point without affecting the physics here. All other parameters are the same as in Eqs. (1) and (3). When $\alpha = 0.5 > 0$, the dispersion of the longitudinal mode tilts downward generating a type-I Weyl point. [Fig. 3(a)]  Whereas when $\alpha = -0.1 < 0$, the dispersion tilts upward, generating a type-II Weyl points [Fig. 3(c)] Figs. 3(b) and 3(d) show the band dispersion in the $k_x - k_y$ plane at $k_z = k_c$ for $\alpha = 0.5$ and $\alpha = -0.1$, respectively. They both possess conical structures around the crossing points between band 1 and band 2. Dispersion along other directions can be found in Supplemental Material Fig. S1. We also note that while the type of the Weyl point changes, the charge of a Weyl point does not change as $\alpha$ varies.

**Physical Realization**

As a physical realization of the dielectric function as discussed above, we consider the structure shown in Fig. 4(a) consisting of a square lattice array of metal wires each forming an elliptical helix. The effective medium theory of such a system was considered in ref. [27]. (The details of the effective medium theory are provided in the Supplemental Material S-IV). The axis of the metallic wire helix is along the z-axis. The wires provide both an effective plasmonic response as well as nonlocality for $\varepsilon_z$. The elliptical shape of the helix provides an anisotropic dielectric response in the x-y plane. Such a system in fact has magnetic response as well with an anisotropic permeability tensor, [27] which however doesn't affect the existence of a Weyl point in this system. (see Supplemental Material S-I). In Figs. 4(b) and 4(c), we provide the band structures of such an elliptical helix array. Fig. 4(b) shows the dispersion along the $z$ direction, which consists of two quasi-transverse modes (red) and one quasi-longitudinal mode (blue line). The quasi-longitudinal mode tilts upward and hence the Weyl points here are of type-II. Fig. 4(c) shows the dispersion along the $x$ and $y$ directions around the crossing point between band 1 and band 3. The band dispersions are linear along all three directions around the crossing point. With the parameters chosen for the structure in Fig. 4, the frequency of the Weyl points here are around 0.14 $c/a$, where $a$ is the lattice constant and $c$ is the speed of light in vacuum, and hence the Weyl points lie in the homogenization limit where the meta-material can be well described by homogenous effective parameters.

In this class of system consisting of a square array of helical wires, Weyl points can in fact be found outside the low-ferquency regime as well. For example, suppose we use a helix that is circular rather than elliptical. From symmetry for the effective parameters we have $\varepsilon_x = \varepsilon_y$, and hence the two quasi-transverse modes are degenerate as can be seen from Eq. (2). On the other hand, the continuous screw symmetry requires that one of the quasi-transverse modes should be degenerate with the quasi-longitudinal mode at the Brillouin zone boundary. Hence away from the low-frequency regime, the degeneracy between these two quasi-transverse modes will be lifted [24] in which case a Weyl point can be found in the absence of in-plane anisotropy, but outside the low-frequency regime.

**Summary and outlook**

Our work here points to the possibility for the exploration of Weyl point physics in reciprocal meta-material systems. The relatively simple design of the structures can certainly be achieved in the microwave frequency range. Moreover, chiral meta-material has been constructed in the mid infra-red wavelength range where the properties of metals can be reasonably approximated by the perfect electric conductor model that we assume here [28,29], thus the physics here can be realized in the optical frequency regime as well. Given the importance of meta-material in both fundamental physics and in device applications, the introduction of Weyl point physics into meta-material will significantly enrich the fields of topological physics as well as metamaterials, leading to new potentials for the manipulation of electromagnetic and optical waves.

This work is supported by the U. S. Air Force of Scientific Research (Grant No. FA9550-12-1-0471).

Q. L. also acknowledges the support of a Stanford Graduate Fellowship.

# Figures

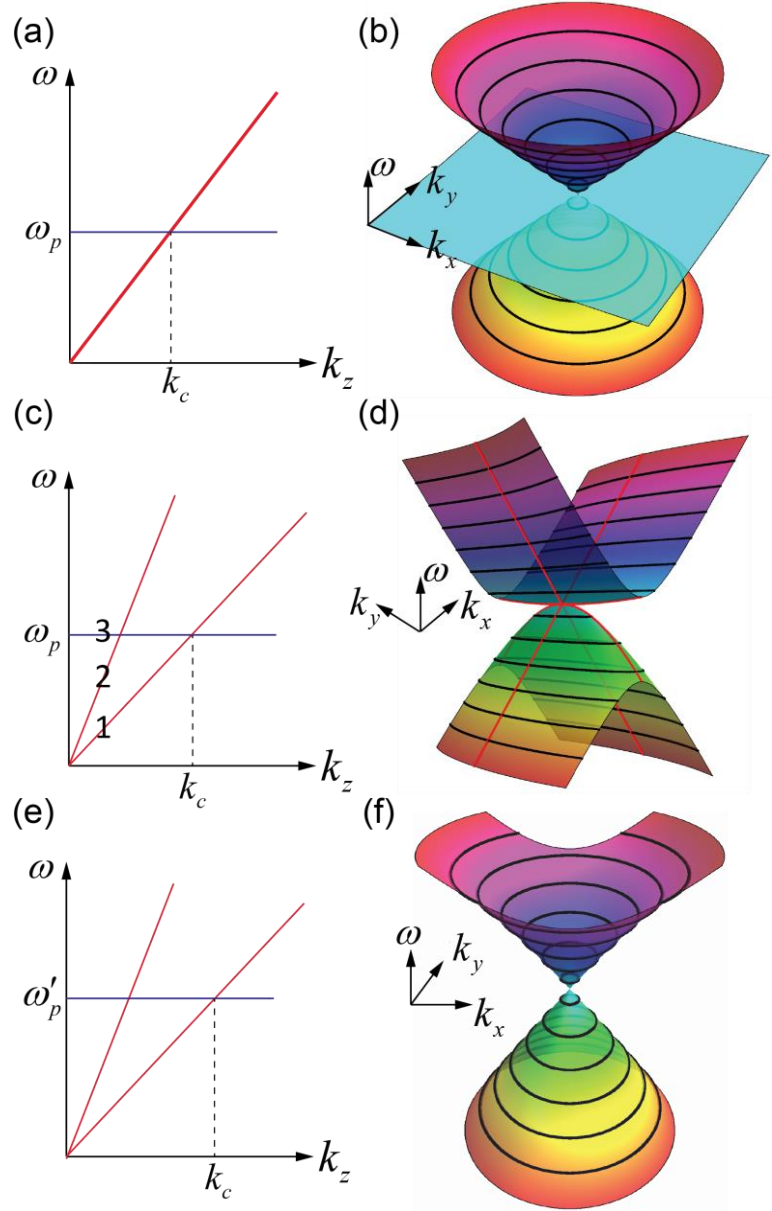

FIG. 1 (color online). (a) The dispersion along the $z$ direction of a system with the relative permittivity given by $\vec{\varepsilon} = \text{diag}(\varepsilon_t, \varepsilon_t, \varepsilon_z)$ where $\varepsilon_z = 1 - \omega^2/\omega_p^2$. The bold red line represents the dispersions of two degenerate transverse modes, and the blue horizontal line represents the dispersion of the longitudinal mode. $k_c$ labels the value of $k_z$ at the crossing point. (b) The dispersion in the $k_x - k_y$ plane at $k_z = k_c$. (c) When anisotropy is introduced and $\vec{\varepsilon} = \text{diag}(\varepsilon_x, \varepsilon_y, \varepsilon_z)$ where $\varepsilon_x \neq \varepsilon_y$, the degeneracy between two transverse modes in (a) splits. (d)

The dispersion around one of the crossing points in (c) at $k_z = k_c$. The dispersion is linear along the *x* direction and quadratic along the *y* direction. (e) The introducing of chiral coupling along the z direction shifts the frequency of the longitudinal mode. (f) The dispersion around the crossing point in (e) at $k_z = k_c$. (e) and (f) show that the dispersion is linear along all the directions indicating the crossing point is a Weyl point. Band numbers are labeled in (c). The parameters used are $W_p = 1$, $\varepsilon_z = 1 - \omega_p^2/\omega^2$ and $\varepsilon_t = 1.8$ for (b); $W_p = 1$, $\varepsilon_z = 1 - \omega_p^2/\omega^2$, $\varepsilon_x = 2$ and $\varepsilon_y = 1.7$ for (d); $W_p' = 1$, $\varepsilon_x = 2$, $\varepsilon_y = 1.7$, $\varepsilon_z = 1 - \omega_p'^2/\omega^2 + \gamma_{zz}^2$, and $\gamma_{zz} = 0.8$ for (f).

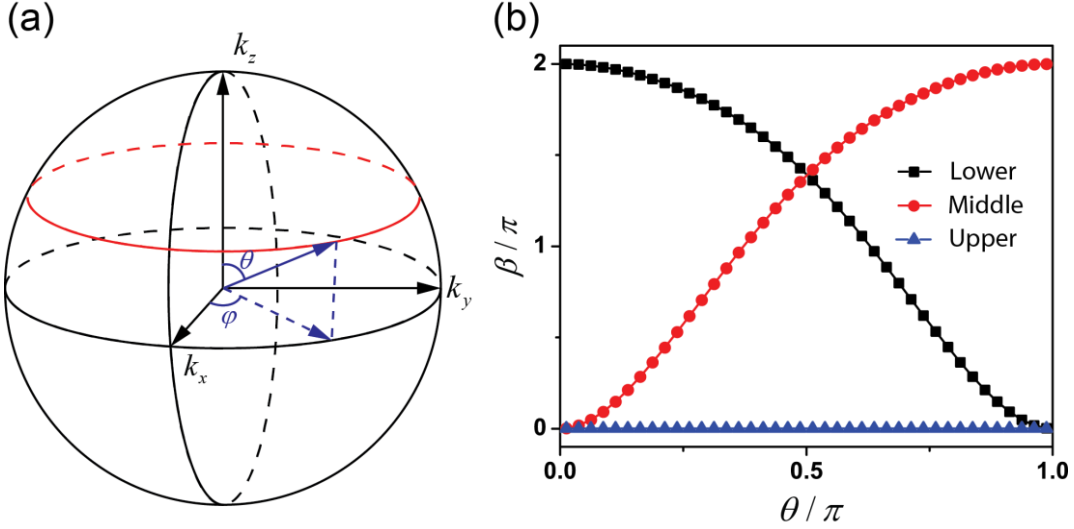

FIG. 2 (color online). (a) Schematic diagram illustrating the method for calculating the charge of a Weyl point. The sphere is centered at the Weyl point in the momentum space, and Berry phase is calculated along the red loop with fixed $\theta$. (b) Berry phases as functions of $\theta$ for these three bands shown in Fig. 1(e). The Berry phase of the lower (middle) band changes continuously from $2\pi$ to $0$ ($0$ to $2\pi$) indicating that the charge of the Weyl point between the lower band and the middle band is $-1$. In contrast, there is no Weyl point on the upper band inside the sphere. The parameters used here are given by $\omega_p'/\omega_p = 1$, $\varepsilon_x = 2$, $\varepsilon_y = 1.7$, $\varepsilon_z = 1 - \omega_p'^2/\omega^2 + \gamma_{zz}^2$, and $\gamma_{zz} = 0.8$. The center of the sphere is set at $k_x = k_y = 0$ and $k_z = k_c = \sqrt{2}\omega_p'/c$ and the radius of the sphere is $0.001 k_c$.

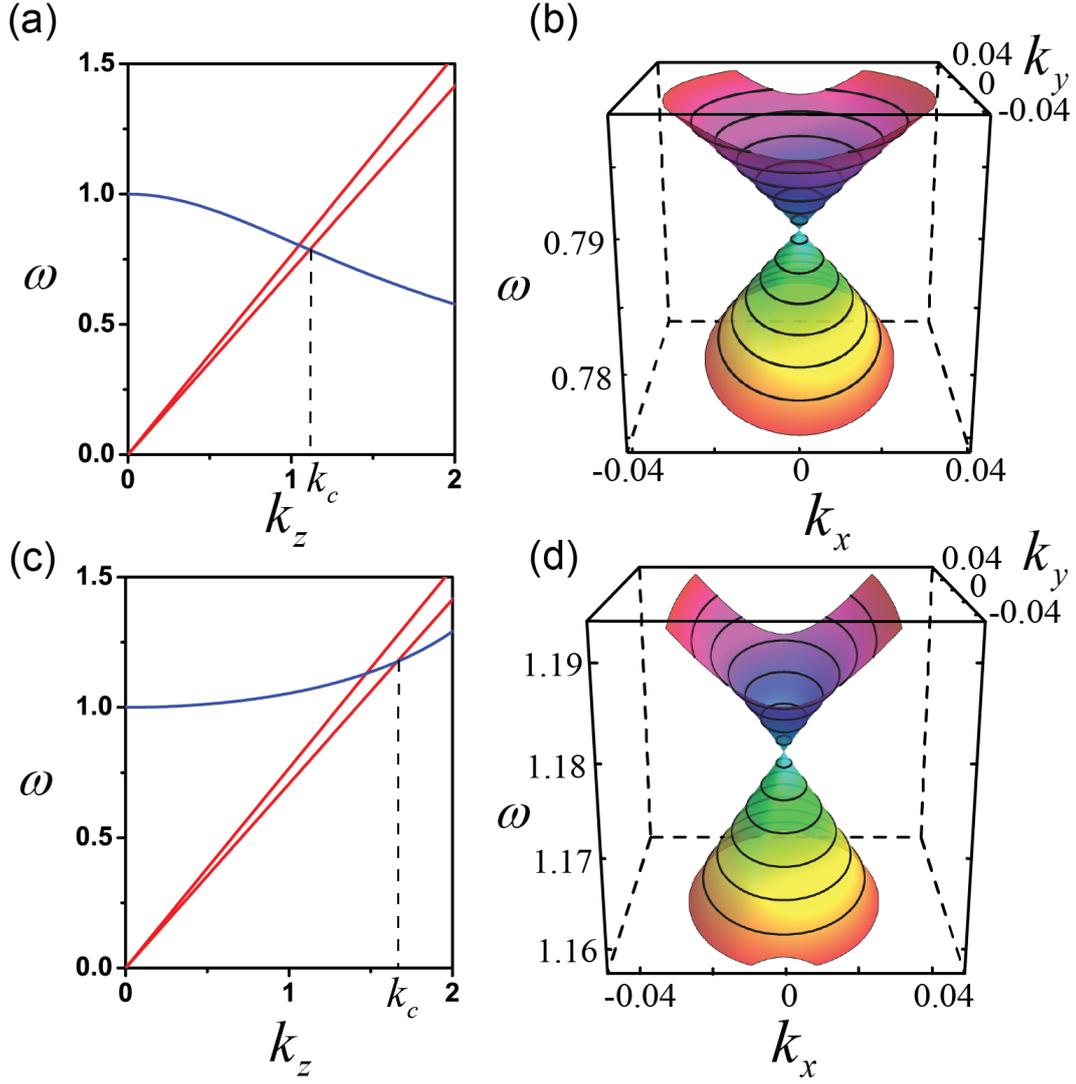

FIG. 3 (color online). (a) and (c) show the dispersion along the $z$ direction, where the red lines and blue line correspond to the transverse modes and longitudinal mode, respectively. (b) and (d) show the dispersions in the $k_x - k_y$ plane at $k_z = k_c$. Both possess conical dispersion. The parameters used here are $\omega'_p = 1$, $\varepsilon_x = 2$, $\varepsilon_y = 1.7$ and $\varepsilon_z = 1 - \omega'^2_p/\omega^2 + \gamma^2_{zz} + \alpha k^2_z$. $\alpha = 0.5$, $\gamma_{zz} = 1.0$ for (a), (b) and $\alpha = -0.1$, $\gamma_{zz} = 0.71$ for (c) and (d). The Weyl points is of type-I when $\alpha > 0$ and type-II when $\alpha < 0$. Here $\omega$ is in unit of $\omega'_p$, and the corresponding wave vectors are in unit of $\omega'_p/c$, where $c$ is the speed of light in the vacuum.

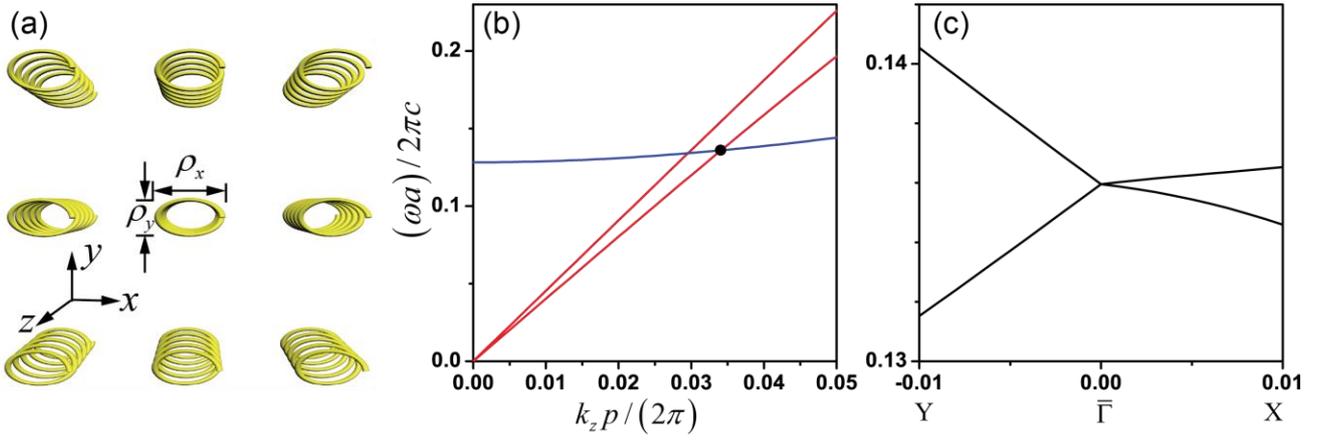

FIG. 4 (color online). (a) The structure consists of a square array of metal wires in the shape of elliptical helix. The use of elliptical helix results in anisotropy in the *x-y* plane. The length of one pitch, the radius of the metallic wire, the long axis and the short axis of the helix are given by $p=0.2a$, $\delta=0.03a$, $\rho_x=0.2a$ and $\rho_y=0.1a$, respectively, where $a$ is the lattice constant. (b) The band structure along the *z* direction. (c) The band structure along the *x* and *y* directions around the crossing point with larger $k_z$ as indicated by a solid black dot in (b). The bands are linear along all the directions indicating that it is a Weyl point. This calculation is performed with Comsol 5.1 and the metal is assumed to be a perfect electric conductor.